\documentclass[aps,prb,twocolumn,superscriptaddress,showpacs]{revtex4}
\usepackage{natbib}
\usepackage{graphicx}
\usepackage{amsmath}
\usepackage{bm}
\usepackage{color,soul}
\usepackage{hyperref}
\usepackage{float}
\usepackage{csquotes}
\usepackage{color,soul}
\hypersetup{colorlinks=true, urlcolor=blue, citecolor=blue}
\usepackage{braket}

\begin{document}
\title{Electron confinement in chain-doped TMDs: A platform for spin-orbit coupled 1D physics}

\author{Mayank Gupta}
\email{Equal Contributors}
\affiliation{Condensed Matter Theory and Computational Lab, Department of Physics, IIT Madras, Chennai 600036, India}
\affiliation{Center for Atomistic Modelling and Materials Design, IIT Madras, Chennai 600036, India}
\author{Amit Chauhan}
\email{Equal Contributors}
\affiliation{Condensed Matter Theory and Computational Lab, Department of Physics, IIT Madras, Chennai 600036, India}
\affiliation{Center for Atomistic Modelling and Materials Design, IIT Madras, Chennai 600036, India}
\author{S. Satpathy}
\email{SatpathyS@missouri.edu}
\affiliation{Condensed Matter Theory and Computational Lab, Department of Physics, IIT Madras, Chennai 600036, India}
\affiliation{Center for Atomistic Modelling and Materials Design, IIT Madras, Chennai 600036, India}
\affiliation{Department of Physics \& Astronomy, University of Missouri, Columbia, MO 65211, USA}   
\author{B. R. K. Nanda}
\email{nandab@iitm.ac.in}
\affiliation{Condensed Matter Theory and Computational Lab, Department of Physics, IIT Madras, Chennai 600036, India}
\affiliation{Center for Atomistic Modelling and Materials Design, IIT Madras, Chennai 600036, India}

\begin{abstract}
The state-of-the-art defect engineering techniques have paved the way to realize novel quantum phases out of pristine materials. Here, through density-functional calculations and model studies, we show that the chain-doped monolayer transition metal dichalcogenides (TMDs), where M atoms on a single the zigzag chains are replaced by a higher-valence transition-metal element M$^\prime$ (MX$_2$/M$^\prime$), exhibit one-dimensional (1D) bands. These 1D bands, occurring in the fundamental gap of the pristine material,  are dispersive along the doped chain but are strongly confined along the lateral direction. This confinement occurs as the bare potential of the dopant chain formed by the positively charged M$^\prime$ ions resembles the potential well of a uniformly charged wire.  
These bands could show novel 1D physics, including a new type of Tomonaga-Luttinger liquid behavior, multi-orbital Mott insulator physics, and an unusual optical absorption, due to the simultaneous presence of the spin-orbit coupling, strong correlation, multiple orbitals, Rashba spin splitting, and broken symmetry. For the half-filled 1D bands, we find, quite surprisingly, a broadening of the 1D bands due to correlation, as opposed to the expected band narrowing. This is interpreted to be due to  multiple orbitals forming the single Hubbard band at different points of the Brillouin zone. Furthermore, due to the presence of an intrinsic electric field along the lateral direction, the 1D bands are Rashba spin-split and provide a new mechanism for tuning the valley dependent optical transitions.
\end{abstract}  
\date{\today}					
\maketitle

\section {Introduction} 
Successful synthesis of atomistically controlled Van der Waals layered materials in the form of transition metal chalcogenides (TMDs) has given rise to a wide range of mesoscopic non-trivial quantum phases. 
These include proximity of $p$-wave superconductivity and charge density wave as in NbSe$_2$ \cite{nature-Xi,Prx-darshana,jpc-REVOLINSKY}, topological Weyl semimetallic nature and large magnetoresistance as in WTe$_2$ \cite{ncomm-Li,nature-Ali,prl-Pletikosi}, and the exotic orbital and quantum spin Hall effects \cite{science-QSHE,Prb-sashi1,Prb-sashi2}. 

One of the emerging areas of research on two- dimensional (2D) TMDs is to further reduce the dimensionality and explore sub-nanoscale quantum physics. For example, the Moir\`e bilayers of TMDs and their heterostructures allow twist angle controlled resonant effects to engineer exciton band structure \cite{Zhang2020, Tran2019}. The lateral superlattices and nanoribbons are synthesized out of TMDs to produce edge and interfacial states \cite{PrbUlloa, LL1-2022, nanoscaleZou, shantanu, JoilePRX, JPCM_Ulloa}. A recent study has proposed a typical Moir\`e lattice out of WTe$_2$ where an overarching periodicity creates a 1D lattice for electrons residing in collective eigenstates which give rise to rarely observed exotic quantum states of Tomonaga Luttinger liquid (TLL) behavior \cite{LL1-2022}. The ribbon edge states in 1-T$^{\prime}$-WTe$_2$ exhibiting TLL behavior is a new experimental result in this direction \cite{shantanu}. The TLL state is also experimentally observed in MoS$_2$  by creating mirror twin boundaries \cite{JoilePRX}. Electron correlation is an important missing factor in TMDs, and therefore, correlation driven exotic quantum phases in the area of magnetism and superconductivity are less evident in this class of compounds. To achieve the correlated electron phases, narrow bands in the vicinity of Fermi energy need to be created, and one of the ways to make it possible is to confine the electron motion by reducing the dimensionality.

Unlike the semimetallic 1T$^\prime$ phase\cite{nature-Keum}, the 2H phase of TMDs has wide band gaps and hence novel quantum phases and transport can be envisaged in them by inducing mid-gap states of lower dimensions. 
\begin{figure}
\includegraphics[scale=0.34]{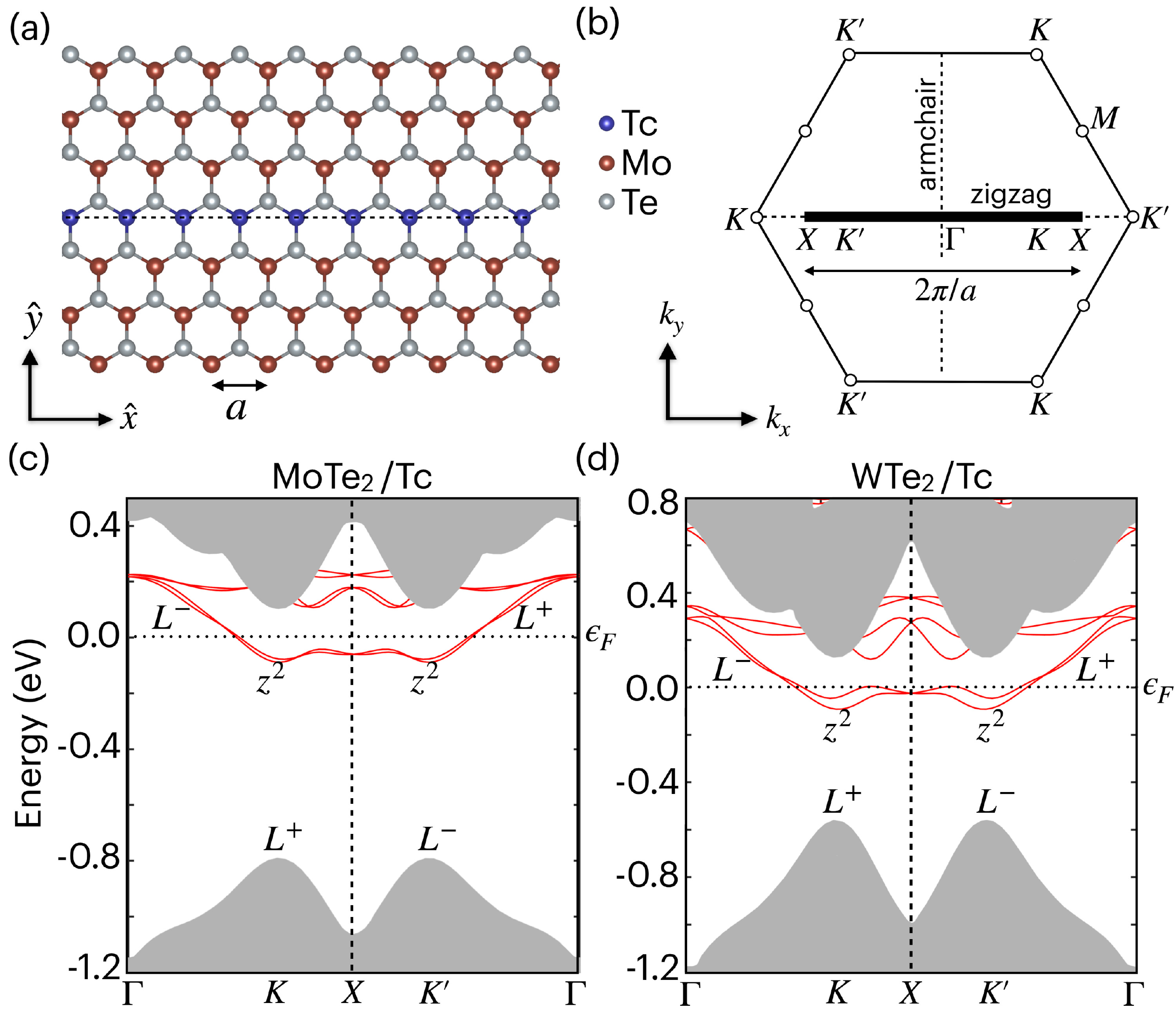}
\caption{(a) Top view of the chain-doped MX$_2$/M$^\prime$ compound,
where a single zigzag chain of the pristine MX$_2$ structure is replaced with an M$^\prime$X$_2$ chain.  
(b) The reduced 1D Brillouin zone  (line extending between $-\pi/a < k_x < \pi/a$) for the chain-doped structure and its relation to the original 2D Brillouin zone (hexagon).
All three valley points $K$  of the hexagonal BZ fall onto the same point, marked by $K$, on the 1D BZ, and the same happens for the $K^\prime$ points.  The hexagonal zone collapses vertically onto the $k_x$ axis, and the points lying outside the 1D BZ in the process, are brought inside it via the reciprocal lattice translation of $2\pi/a$.
The ``zigzag/armchair" labels in the figure indicate the orientation of the BZ with respect to the crystalline directions in real space. (c) and (d) The DFT+SOC band structure of the pristine MoTe$_2$ and WTe$_2$ (shaded grey), projected into the 1D BZ. The red lines indicate the defect bands introduced by the Tc doped chain in the forbidden region. These bands represent 1D propagating states along the chain, while they are confined in the lateral direction. The defect bands are dominated by the orbital characters of Tc. The small splitting of the otherwise degenerate defect bands is because of the Rashba SOC due to a non-zero lateral electric field (see text).
 }
\label{Fig1}
\end{figure}
With growing interest in doped 2D TMDs, there are experimental proposals on tunable doping mechanisms in these systems \cite{Gwangwoo}. Recently, Lin $\textit{et.  al.}$ \cite{Lin-2021} reported excellent controllability for substitutional doping of the foreign atoms in 2D TMDs through low energy ion implantation techniques such as site-selective laser-assisted chemical vapor doping \cite{Kim-2016}. Furthermore, a very recent experimental work demonstrates a controlled doping strategy for TMDs based on a dislocation climb mechanism  \cite{Lin2-2021}. In this study, the authors were successful in forming highly doped nanostripes of Ti, V, Cr, and Fe atoms in WSe$_2$ and WS$_2$ monolayers.

In this paper, we have engineered 1D quantum states out of the semiconducting 2H phase of MoX$_2$ and WX$_2$ monolayers with X being a chalcogen. This is achieved by replacing a single chain of Mo or W along the zigzag direction with an element  M$^\prime$ with one extra valence electron as shown in Fig. \ref{Fig1}. We find these chain-doped systems, henceforth represented as MX$_2$/M$^\prime$ to be dynamically stable. The 1D bands, depending on several other factors, build a perfect platform to induce non-trivial low-dimensional quantum phases, which may include Peierls distortion \cite{Barborini2022}, topological magnons \cite{Su2018}, charge density waves \cite{Hofmann2019}, TLL \cite{sashi3-2006}, and 1D magnetism.

In the case of MTe$_2$/Tc and MTe$_2$/Re, we find that the weakly SOC driven degenerate half-filled 1D bands running along the chain make them ideal candidates for exhibiting TLL phenomena. In the magnetic phase, the effect of strong correlation can produce Mott insulating states by breaking the half-filled 1D bands to lower and upper Hubbard subbands with a gap in between. However, a new phenomenon emerges where the onsite Coulomb repulsion instead of localizing, delocalizes the lower Hubbard subbands. Upon practical realization, it can give rise to an unusual 1D quantum transport. The dopant chain breaks the reflection symmetry to introduce an intrinsic electric field along the lateral direction. This in turn, makes the 1D bands Rashba spin-split and introduces valley dependent optical transition in the system.
\begin{figure}[!ht]
\includegraphics[scale=0.4]{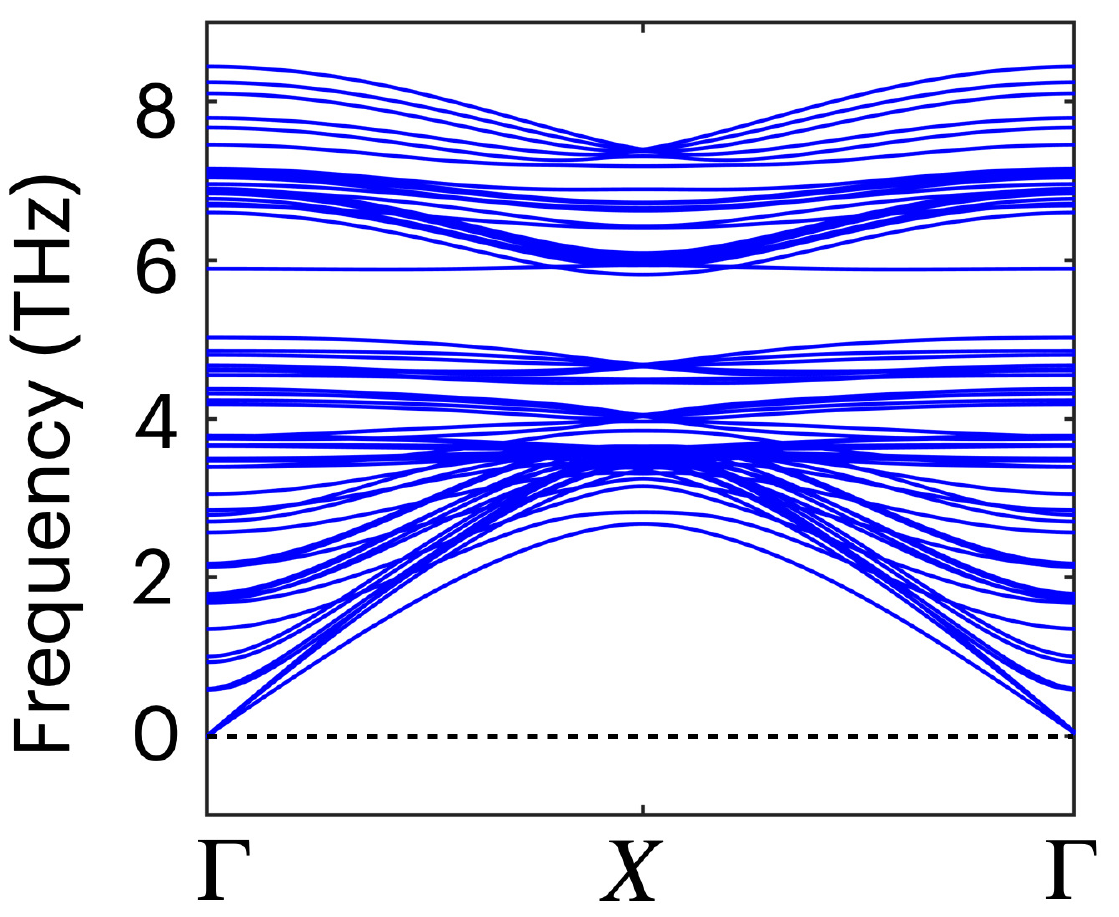}
 \caption{The phonon frequencies of the chain-doped MoTe$_2$/T$_c$ (Mo$_6$Tc$_1$Te$_{14}$). The force constants obtained from the DFPT method are taken into account through the phonopy code  \cite{phonopy} as implemented in VASP. The absence of imaginary modes implies the dynamical stability of the chain-doped structure.}
\label{Fig2}
\end{figure}

\section{Structural and Computational Details} The prototype representation of the chain-doped monolayer structures (MX$_2$/M$^\prime=$M$_{n-1}$M$^\prime_1$X$_{2n}$) is shown in Fig. \ref{Fig1} (a). In this study, the supercell approach is adopted with $n$ = 13, which is found to be sufficient to induce the 1D defect bands. The phonon band structures of such systems do not show imaginary frequencies (see Fig. \ref{Fig2}), suggesting dynamical stability. With chain doping along the zigzag direction, the 2D Brillouin zone (BZ) reduces to a 1D BZ as shown by a thick black strip in Fig. \ref{Fig1} (b). The high symmetry points of the 2D BZ are projected onto the reduced BZ, which helps us later in discussing the resonance and bound states. 
\begin{figure}
\centering
\includegraphics[scale=0.36]{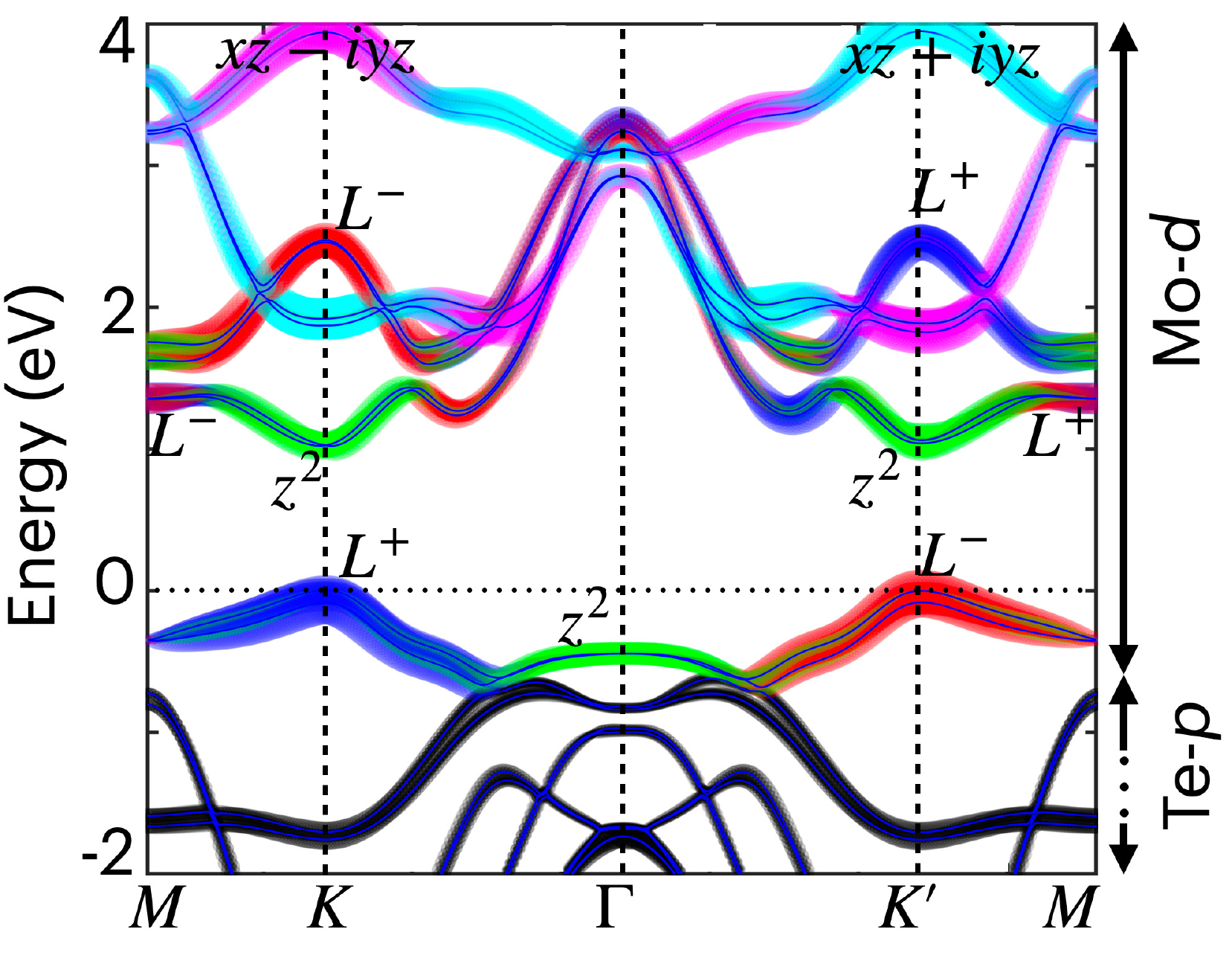} 
 \caption{The orbital resolved band structure of the monolayer MoTe$_2$ obtained using Wannier90 \cite{MOSTOFI2008685}. The valence band maximum occurring at the valley points, $K$ and $K^\prime$, are dominated by the angular momentum orbitals L$^\pm = (x^2-y^2 \pm ixy)$, while the conduction band minima are formed by the $z^2$ orbital, all belonging to the Mo atom. The spectrum below the top valence band is formed by the Te-$p$ states. }
\label{Fig3}
\end{figure}
\begin{figure*}
\includegraphics[scale=0.55]{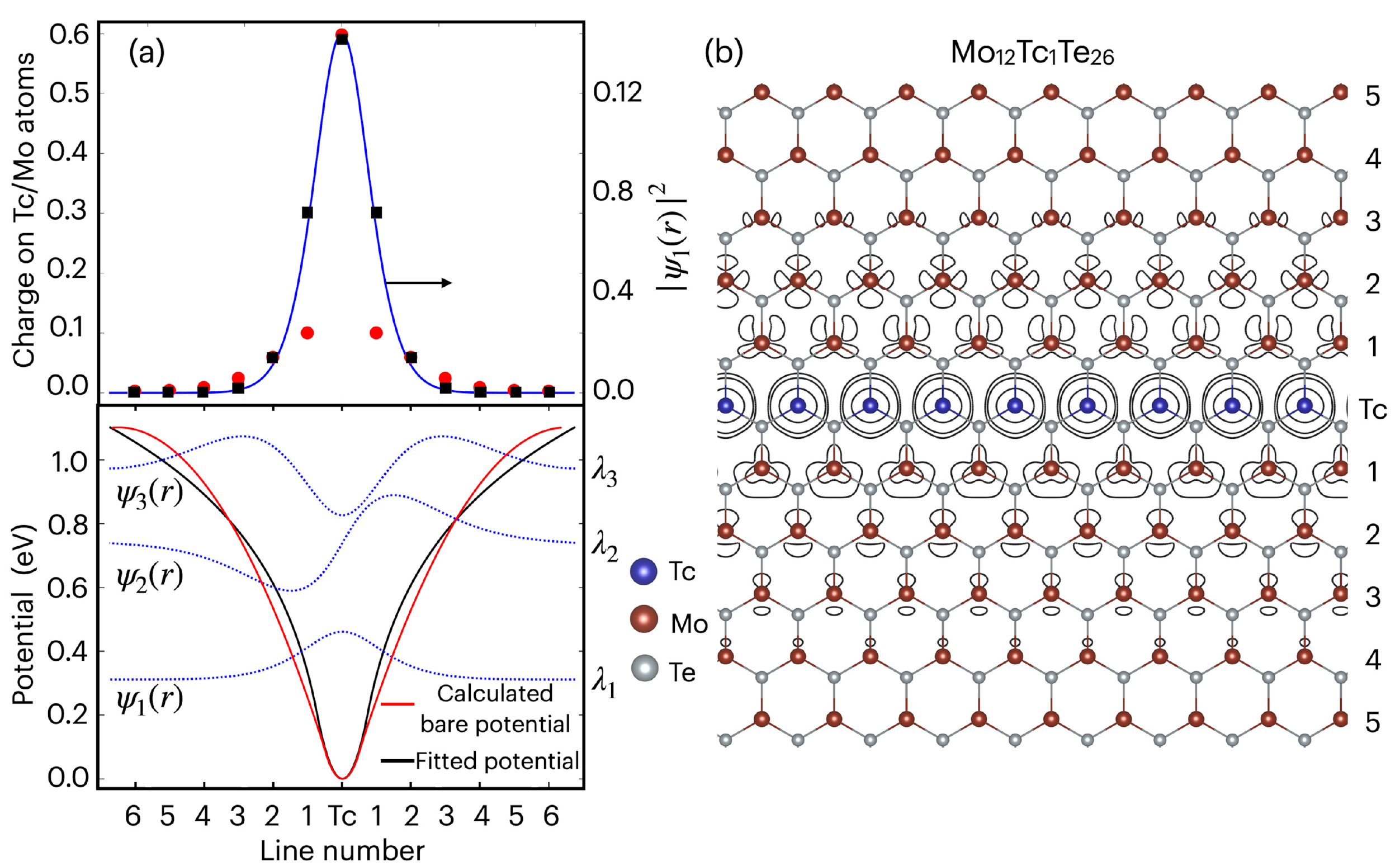}
 \caption{The confinement of the chain-doped bands along the lateral direction. (a) Lower panel: The cell-averaged bare potential (red solid line) and the model potential (black solid line) as per Eq. 1. The ground and the first two excited states wave functions ($\psi_1$, $\psi_2$, and $\psi_3$) are sketched in blue dotted lines. These are obtained by numerically solving the Schr\"odinger equation for the model potential. Upper panel: The spread of the extra valence electron in this potential. The DFT obtained values are shown in red circles. The cell average of the ground state charge density ($|\psi_1(r)|^2$) is shown in black squares. (b) Charge-density contours (isosurface value = 0.0001 $e$/$\AA^3$) of the partially occupied defect bands as obtained by integrating from the bottom of the defect bands to the Fermi energy (see the red bands in Fig. \ref{Fig1}).}
 \label{Fig4}
\end{figure*}

The density functional theory (DFT) calculations are carried out on optimized M$_{n-1}$M$^\prime_1$X$_{2n}$ structures using the pseudopotential based projector-augmented wave (PAW) method \cite{Blochl1994,Kresse1999} within the framework of PBE-GGA exchange-correlation functional as implemented in Vienna ab initio simulation package (VASP) \cite{Kresse1996}. The plane wave energy cut-off of 400 eV and a 1$\times$8$\times$2 $\Gamma$-centered k-mesh is used for BZ integration. The Hubbard $U$ formalism is adopted to study the correlation effect arising due to localized defect states. The $U$ values are obtained using the linear response theory \cite{Cococcioni2005}. The cell averaged bare potential is calculated using the QUANTUM ESPRESSO simulation package \cite{Giannozzi_2009}.

\section{Formation of 1D bands} While a range of chain-doped configurations are investigated, here, we will discuss the electronic structure of MoTe$_2$/T$_c$ and WTe$_2$/T$_c$ as prototypes. However, it is useful to first provide a brief overview of the electronic structure of the pristine TMDs so that the formation of the dopant states can be better understood. For this purpose, in Fig. \ref{Fig3}, we present the band structure of monolayer 2H-MoTe$_2$ \cite{Prl-xiao}. The electronic properties of the TMDs have been widely studied \cite{Prl-xiao, Manzeli2017, ChemicalRev2020}. The formation of bands can be described with two step chemical bonding. In the first step, the nearest neighbor Mo-$d$ -- Te-$p$ interactions give rise to lower-lying Te-$p$ dominated bands and upper-lying Mo-$d$ dominated bands. Driven by the trigonal prismatic crystal field, the latter is further split into three groups: A$_1^\prime$(${z^2}$), E$^\prime$(${xy}$, $x^2-y^2$) and E$^{\prime\prime}$(${xz}$, ${yz}$). The second step involves second neighbor interactions where in the monolayer limit the reflection symmetry along $\hat{z}$ direction permits hybridization among the A$_1^\prime$ and E$^\prime$ orbitals  to create a band gap. The valley points $K$ and $K^\prime$ host both the valence band maximum (VBM) and the conduction band minimum (CBM). The VBM at $K$ and $K^\prime$ are found to be formed by $x^2-y^2 + ixy$ ($L^+$) and $x^2-y^2 - ixy$ ($L^-$) orbitals respectively, giving rise to opposite orbital moments \cite{Prb-sashi1} while the CBM is formed by the $z^2$ character and hence with zero orbital moments \cite{Prb-sashi1, Prl-xiao}. The role of the spin-orbit coupling (SOC) in this compound is restricted to splitting the bands dominated by $L^+$ and $L^-$ by a few meV without perturbing the broad band structure. To produce unique quantum transport phenomena, the mid-gap states with varying characters can be generated in these systems through hole or electron doping.

The band structures of the chain-doped systems MoTe$_2$/Tc and WTe$_2$/Tc are respectively shown in Figs. \ref{Fig1} (c) and (d). The gray shaded region represents the bands of pristine MoTe$_2$/(WTe$_2$) projected along the k$_y$ = 0 (see Fig. \ref{Fig1} (c)/(d) ). The red bands belong to the chain-doped compounds. Most of them overlap with the  bands of the parent compound, but the rest form defect bands, creating either bound states by lying in the forbidden region or resonating states by overlapping with the bulk bands. The defect bands lying in the vicinity of the Fermi level ($\epsilon_F$) are of significant importance as they can introduce new transport behavior in the system. Our orbital projection analysis indeed shows that these bands are formed by the $xy$, $x^2-y^2$, and $z^2$ orbitals of the Tc chain. Furthermore, the defect bands are dispersive along the chain direction while remaining bound perpendicular to it and thereby a platform for 1D quantum physics is created.

The basic electronic configuration enables us to explain the formation of the 1D quantum state. In MoTe$_2$ and WTe$_2$, Mo$^{4+}$ and W$^{4+}$ has $d^2$ configuration while Tc offers one additional electron with $d^3$ electronic configuration. Therefore, when a chain of Tc is placed in the Mo/W matrix the former adopts a $d^2$ + $d^1$ electronic configuration. The $d^2$ primarily contributes to the bulk while the Tc-$d^1$ is responsible for forming the defect bands.

By giving away the additional electron ($d^1$), the Tc chain behaves like a positively charged wire of radius $R_0$. From Gauss's law, the potential inside and outside the wire can be expressed as:
\begin{equation}
  V = \left\{
    \begin{array}{ll}
      \frac{\rho}{4\epsilon_r\epsilon_0}r^2, & r < R_0 \\
         \frac{\rho R_0^2}{4\epsilon_0\epsilon_r}\bigg(1 + 2 \log(\frac{r}{R_0})\bigg) , & r > R_0
    \end{array}
  \right.
\end{equation}
where, $\rho$ = $e/\pi R_{0}^2a$ is the charge density of the wire with $a$ being the lattice constant. Based on the earlier theoretical studies the dielectric constant ($\epsilon_r$) is taken to be 20 \cite{dielectric_MoTe2}. We mapped the modeled potential with the cell averaged bare potential obtained from the DFT calculations on a MoTe$_2$/Tc superlattice. There is an excellent agreement capturing both $r^2$ and logarithmic behavior  inside and outside the wire respectively for $R_0$ = 2.6 $\AA$ which is higher than the atomic radii of Tc and lower than the lattice parameter. 

The wave functions of the ground and first two excited states ($\psi_n(r)$) and their corresponding eigenvalues ($\lambda_n$) are shown in Fig. \ref{Fig4} (a) lower panel, which were obtained from the numerical solution to the one-particle Schr\"odinger equation with the potential given in Eq. 1. The eigenstates resemble those of the Airy functions which are the solutions of the Schr\"odinger equation for at linearly varying potential well \cite{Popovic}.
The $|\psi_1(r)|^2$ plotted in a blue solid line in the upper half of Fig. \ref{Fig4} (a) reflects the charge spread away from the Tc chain. For validation, we computed the cell average of $|\psi_1(r)|^2$ (black filled squares) along the direction perpendicular to the chain and compared them with the atom wise contribution of partially occupied lower lying defect band, and values are depicted through the red filled circles and found a very good match among them. The rapid exponential decay of the charge spread is also demonstrated through the logarithmic charge density ($\rho_{DFT}(r)$) contours calculated for the 1D band. The $\rho_{DFT}(r)$ is calculated by integrating the lower lying defect band upto $\epsilon_F$. From Fig. \ref{Fig4} (b), we observe that the spread vanishes after three layers on either side of the Tc chain while accumulating most of the charge on the chain itself. This implies that the defect state is bound laterally and dispersive along the Tc chain.\\
\begin{table}
    \centering
        \caption{Rashba strength (in eV $\cdot$ \AA) and the hopping integrals (in eV) for the 1D defect bands in MX$_2$/M$^\prime$.}
    \begin{tabular}{|c|c|c|c|c|}
\hline
        Compound \hspace{0.1cm}name &$\lambda_R$& $t$ & $t^\prime$ & $t^{\prime\prime}$ \\
         \hline
         MoSe2/Tc & 0.10  & -0.053 & 0.032 & 0.003 \\
         \hline
         MoSe2/Re & 0.76  &  -0.049 & 0.035 & 0.019 \\
         \hline
         MoTe2/Tc & 0.10  & -0.070 & 0.026 & 0.007 \\
         \hline
         MoTe2/Re & 1.00  & -0.077 & 0.030 & 0.019 \\
         \hline
         WS2/Tc &  0.00 & -0.041 & 0.039 & 0.002 \\
         \hline
         WSe2/Tc & 0.20  & -0.057 & 0.044 & 0.000 \\
         \hline
         WTe2/Tc &  0.74   & -0.090 & 0.048 & 0.007 \\ 
         \hline
         WTe2/Re &  0.68 & -0.081 & 0.041 & 0.025 \\
         \hline
    \end{tabular}
    \label{Table-I}
\end{table}
\section{Platform for 1D physics}
\subsection{Rashba SOC and valley-dependent optical transitions}
The partially filled non-degenerate defect bands can be fitted with a tight-binding model on the doped chain along with a Rashba-like term, viz., 
\begin{equation}
    E = (\varepsilon_0 + 2t \cos \ k_x+2t^\prime \cos \ 2k_x+2t^{\prime\prime} \cos \ 3k_x) I
    + \lambda_R (\hat{E}\times\vec{k})\cdot\vec{\sigma},
    \label{TBenergy}
\end{equation}
where the chain runs along $\hat x$, $\varepsilon_0$ is the on-site energy taken to be zero, $t$, $t^\prime$, and $t^{\prime\prime}$ are, respectively, the hopping to the first, second, and the third nearest neighbor. Here, $I$ is 2$\times$2 identity matrix and $\vec{\sigma}$ are Pauli spin matrices. $\lambda_R$ is the Rashba strength.
From the symmetry of the structure (Fig. \ref{Fig4} (b)), an electric field exists in the $y$ direction on the plane, which translates to a magnetic field $\vec B = \vec v \times \vec E/ c^2$ in the electron's rest frame that couples to the spin moment. This leads to the spin-split band structure with a linear dispersion described by the last term in Eq. (\ref{TBenergy}). 

The TB parameters, obtained by fitting to the DFT results, are listed in Table \ref{Table-I} for a number of chain-doped compounds. We note that there is a substantial 2$^{nd}$ neighbor hopping $t^\prime$, but the 3$^{rd}$ neighbor hopping is substantial only for the Re chains and negligible for the Tc chains. The TB bands fitted with DFT for WTe$_2$/Tc is shown in Fig. S10 of supplementary materials (SM). Similar models can also be developed for the Rashba spin-split defect bands that lie in the forbidden regions other than the fundamental gap.\par
\begin{figure}
    \centering
    \includegraphics[scale=0.27]{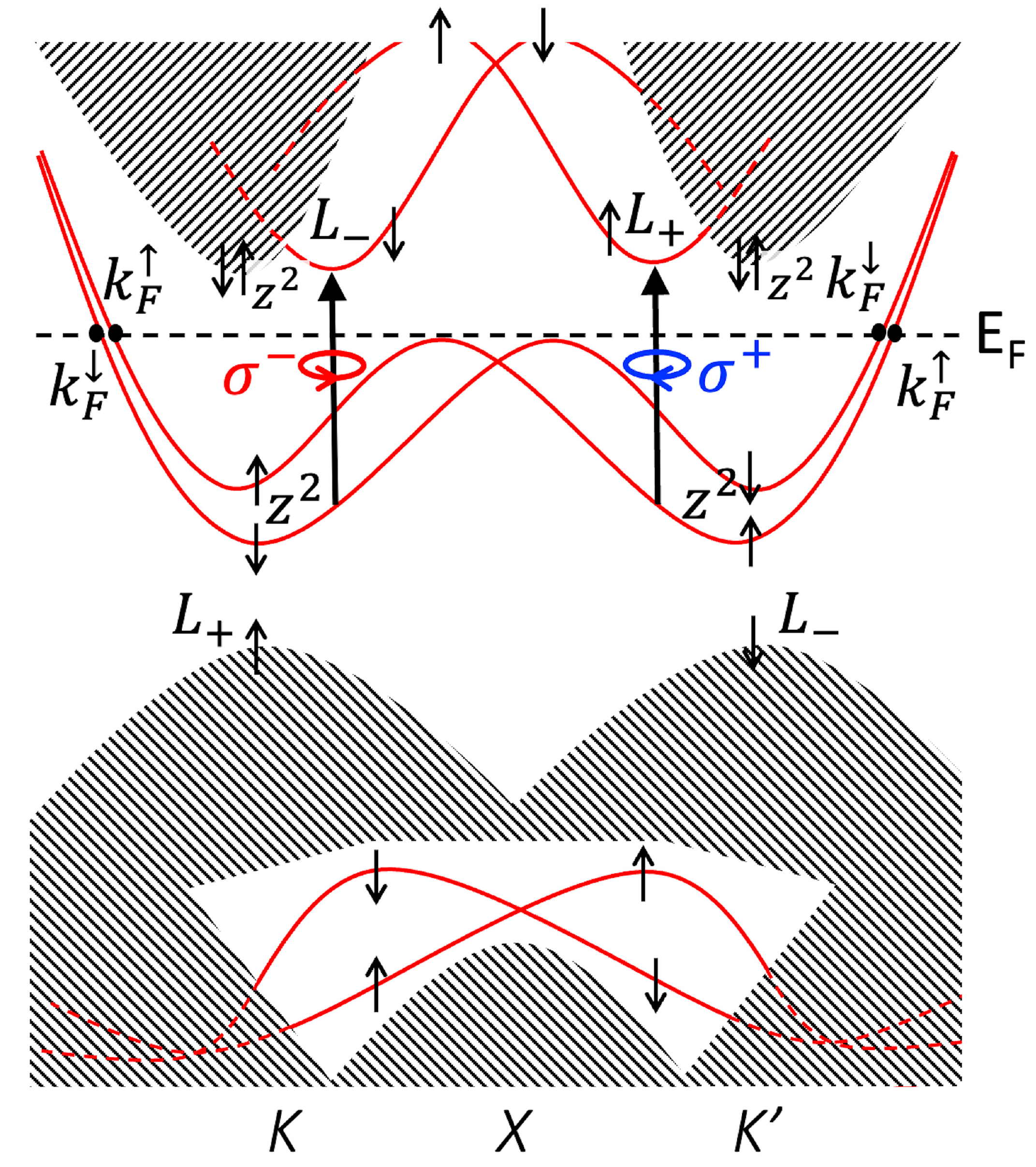}
    \caption{Schematic 1D defect bands indicating the Rashba spin splitting and valley dependent optical absorption. The shaded bands indicate the band structure of the 2D host material, while the red lines indicate the localized 1D defect bands introduced in the band gap of the host. The red dashed lines in the valence band indicate resonance states. Some  dipole-allowed optical transitions for circularly polarized light are indicated by $\sigma^-$ and $\sigma^+$.
  The orbital projected bands of prototype WTe$_2$/Tc are shown in Fig. S11 of SM.}
    \label{Fig5} 
\end{figure}
The schematic band structure along with the Rashba spin splitting is illustrated in Fig. (\ref{Fig5}), which suggests interesting valley dependent optical properties. In the parent MX$_2$ material, circularly polarized light with opposite polarization is absorbed at the two different valleys $K$ and $K^\prime$. Due to the Rashba-like spin splitting, we predict the chain-doped compounds to exhibit additional features in the valley-dependent optical absorption between the bulk to the defect states. The lowest-energy optical transitions at the valley points are forbidden because the lower defect band has spin opposite to that of the bulk valence band edge. Note that the projected bands onto the 1D BZ of the chain-doped compound not only shows the fundamental gap extending through the full BZ, but also gaps that exist at certain regions of the BZ as indicated for the valence bands in Fig. \ref{Fig5}. As also indicated in the figure, localized 1D bands can exist within these gaps, and valley-dependent optical transitions would occur between these states including the localized 1D bands lying in the fundamental gap. To illustrate, the dipole allowed lowest-energy optical transitions ($\sigma^+$ and $\sigma^-$) between partially filled and localized 1D conduction bands in the vicinity of valley points are indicated in the figure. 
\begin{figure}
\includegraphics[scale=0.43]{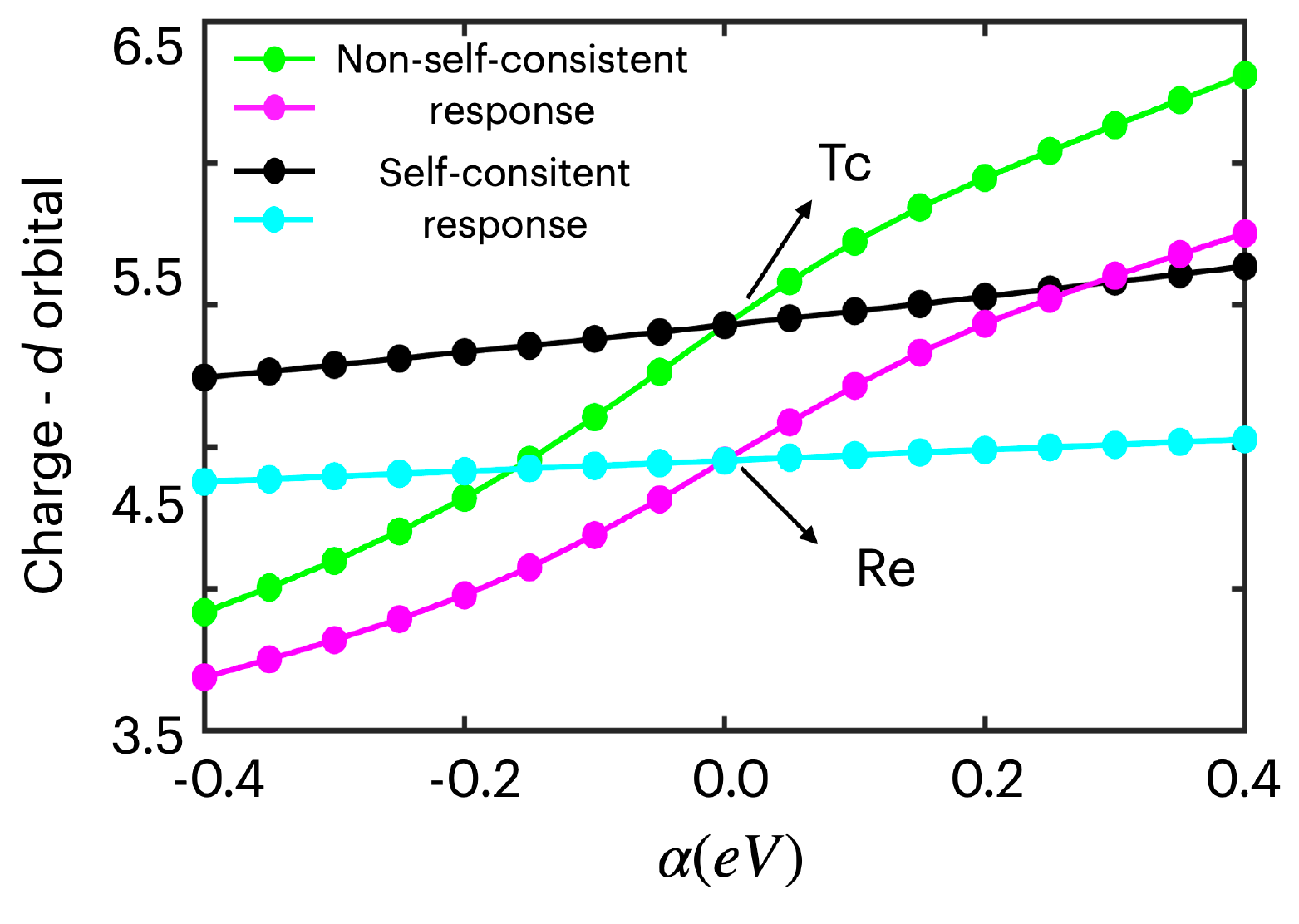}
    \caption{Calculation of the Coulomb repulsion $U_0$ for the metal atom on the doped chain, Tc or Re, using Eq. \ref{Coulomb}. 
    Plotted are the occupation numbers, $n_i^{KS}$ and $n_i$, as a function of the perturbing potential $\alpha$ at the doped metal site, obtained using 1$\times$4 supercell of MoSe$_2$/Tc and MoSe$_2$/Re, i.e. (M$_{n-1}$M$^\prime_1$X$_{2n}$, $n$ = 7)$\times$4, where the perturbing potential was applied to a single Tc or Re atom.}
    \label{Fig7}
\end{figure}
\subsection{Electronic Correlation}
Electron correlation effects are expected to be important for the 1D defect bands, since the bandwidth is quite narrow. A key parameter for characterizing the strength of the correlation effects is the on-site Coulomb repulsion $U$, which we now proceed to compute using the DFT and the linear response approach.

\begin{figure}[!ht]
\includegraphics[scale=0.50]{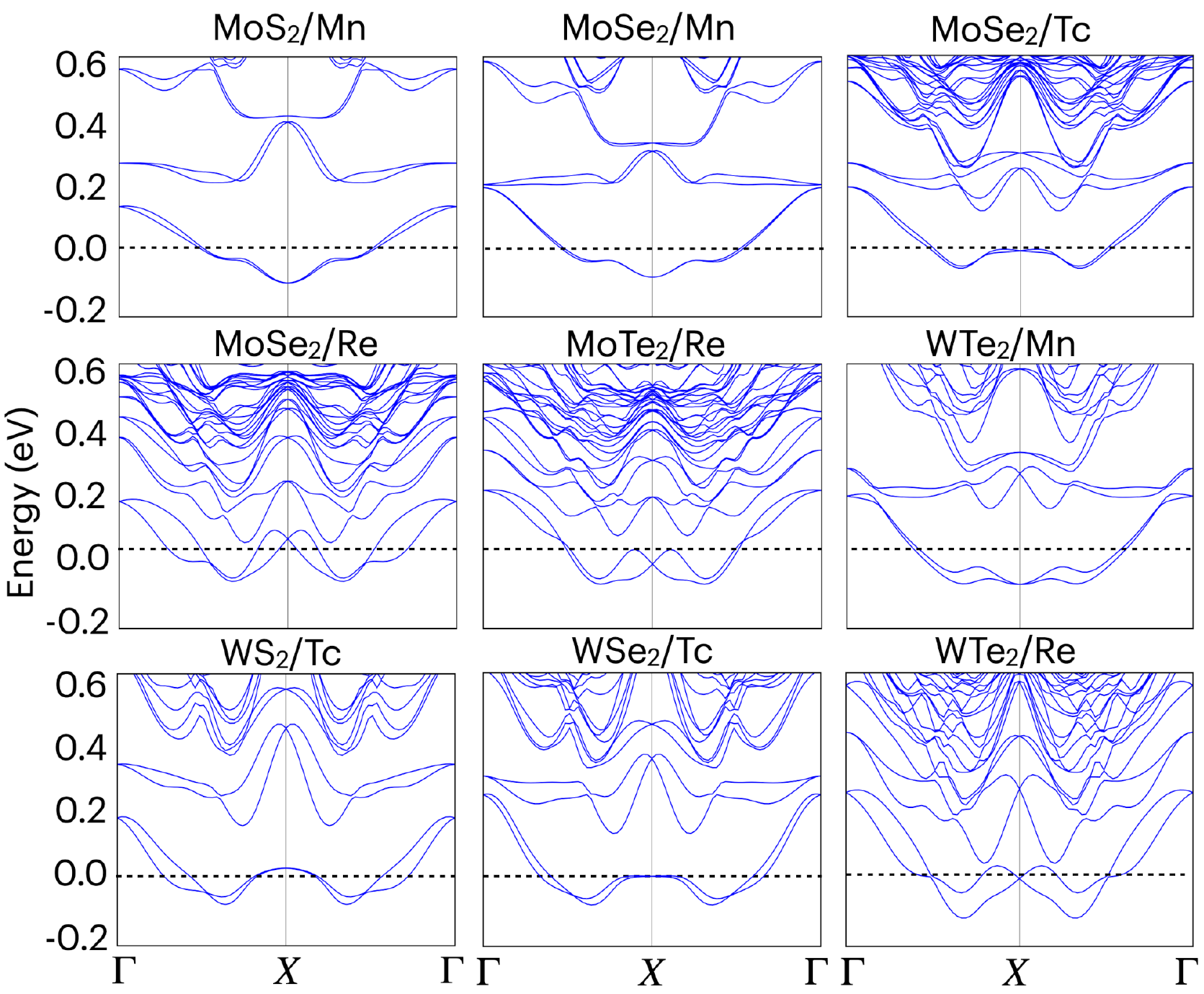}
    \caption{The non-magnetic band structures of several chain-doped compounds. The SOC is included.}
    \label{Fig6}
\end{figure}

In this method \cite{Cococcioni2005}, $U$ is computed by calculating the difference between interacting and non-interacting density response functions:
\begin{eqnarray}
    U = \chi^{-1}_0 - \chi^{-1} 
      =  \Big(\frac{\partial n_i^{KS}}{\partial \alpha_i}\Big)^{-1} - \Big(\frac{\partial  n_i}{\partial \alpha_i}\Big)^{-1},
      \label{Coulomb} 
\end{eqnarray}
where $\alpha_i$ is a perturbative shift in the single particle potential at site $i$, for which $U$ is being computed. Since the bandwidth of the $d$ bands of the host compound MX$_2$ are rather large, and the bands are either occupied or unoccupied, the correlation effects are relatively weak. In contrast, the defect bands are one-dimensional, relatively narrow, and half-filled, so that the correlation effects are expected to be important there. Therefore, we compute $U$ only for the 1D defect bands. To obtain the response functions, the variation in occupation numbers is obtained by performing the DFT calculations in two ways: (i) by allowing the Kohn-Sham potential to adjust self-consistently which optimally screens the perturbation $\alpha_i$ to give $\chi_0$ and (ii) by calculating the Kohn-Sham potential without screening to get $\chi$. The latter is achieved by a single loop, without enforcing self-consistency. The variation of $n_i^{KS}$ and $n_i$ as a function of $\alpha_i$ at the doped metal site (Tc or Re) is shown in Fig. \ref{Fig7} for MoSe$_2$/Tc and MoSe$_2$/Re. The onsite Coulomb repulsion $U_0$ calculated with this procedure is listed for various chain-doped compounds in Table \ref{Table-II}, together with the Fermi velocity $v_F$ and the bandwidth $W$. The spin resolved Fermi velocities $v^\uparrow_F$ and $v^\downarrow_F$ are computed at the Fermi momentum $k^\uparrow_F$ and $k^\downarrow_F$, which is roughly halfway along the $\Gamma - X$ line (see Figs. \ref{Fig1} and \ref{Fig6}), by taking the derivative of the energy $ \hbar v^{\uparrow\downarrow}_F= (\partial E^{\uparrow\downarrow}(k)/\partial k)_{k=k^{\uparrow \downarrow}_F}$. The $k^\uparrow_F$ and $k^\downarrow_F$ differ as the SOC makes the spin-resolved bands non-degenerate. As seen from Table \ref{Table-II}, $U/W >> 1$ for all compounds studied, which would put these materials in the strong correlation limit.
\begin{table}
\caption{Characteristics of the 1D defect band for various compounds. $U$, $W$, and $v_F$ are the onsite Coulomb repulsion, bandwidth, and Fermi velocity, respectively.}
\begin{tabular}[t]{|p{2cm}|c|c|c|c|c|}
\hline
Compound/ &$U$(eV) & W(eV) &$v_{F\uparrow}$(eV\AA)&$v_{F\downarrow}$(eV\AA)\\ 
doping element&&&&\\
\hline
MoTe$_2$/Tc&1.65&0.30&0.67&0.78\\ \hline
MoSe$_2$/Tc&1.63&0.26&0.41&0.50\\ \hline
WTe$_2$/Tc&1.65&0.34&0.50&0.73\\ \hline
WSe$_2$/Tc&1.63&0.31&0.64&0.76\\ \hline
WS$_2$/Tc&1.30&0.20&0.43&0.76\\ \hline
MoSe$_2$/Re&4.99&0.24&0.82&0.58 \\ \hline
MoTe$_2$/Re&4.90&0.27&1.01&0.69 \\ \hline
WTe$_2$/Re&4.96&0.31&1.11&0.73 \\ \hline
\end{tabular}
\label{Table-II}
\end{table}
\subsection{Tomonaga Luttinger-Liquid Physics}
The low-energy behavior of the correlated electrons in 1D is described by the TLL theory, with features generic to many interacting 1D electron systems, such as the spin-charge separation and the anomalous scaling of the correlation functions. \cite{Tomonaga_1950, LL_Luttinger, LL_Haldane, LL_1D} \cite{giamarchi2003quantum}. A few years ago, the TLL theory was extended to include SOC. When SOC is introduced, the complete spin-charge separation is destroyed, resulting in mixed bosonic excitations involving these two degrees of freedom.
  However, the anomalous scaling of the correlation functions remains with modified exponents \cite{moroz_2000, Sedlmayr_2013}. The TLL with SOC has been studied experimentally in the context of quasi-1D systems such as the carbon nanotubes, quantum wires, and 2D electron gases confined to a narrow channel by gate electrodes, but much remains to be done both from theoretical and experimental points of view.

We propose the chain-doped TMDs to be an important class of materials to study the TLL physics. As illustrated for a number of cases in Figs. \ref{Fig1} and \ref{Fig6}, the defect bands in the chain-doped TMDs are spin split at the Fermi energy to a varying degree due to the SOC. For example, the splitting is near zero for MoTe$_2$/Tc, while it is quite large for WS$_2$/Tc.
The 1D defect bands in the chain-doped TMDs are nominally half-filled, for which the TLL behavior would be absent \cite{Note1}. However, departure from the half-filling scenario  can occur naturally due to the presence of impurities and/or applied gate voltage, which may furthermore be used to  alter the spin-splitting at the Fermi energy. Thus the chain-doped TMDs may serve as a rich laboratory for the study of TLL behavior.

The TLL behavior is characterized by several parameters such as the spin/charge velocities for the collective spin and charge excitations and the anomalous dimension $\alpha$ which is a function of onsite Coulomb repulsion and $v_F$. Explicit expressions for these quantities exist in the weak-coupling limit, but for the strong-coupling limit, numerical results have been obtained for certain models. For the 1D Hubbard model, the TLL parameters have been computed for different values of $U/t$ and band filling $n$ \cite{Schulz_1990}. For $U/t \rightarrow \infty$, which is the case for the chain-doped compounds, one gets
$\alpha = 1/8$. When SOC is present, its strength is parameterized by the Fermi velocity difference
$(v_{F\uparrow}-v_{F\downarrow})/(v_{F\uparrow}+v_{F\downarrow})$, and the TLL exponent $\alpha$ is modified \cite{moroz_2000}. The TLL behavior of the 1D Hubbard model with next-nearest-neighbor hopping but without the SOC has been studied using the density-matrix renormalization group \cite{Nishimoto}. For the chain-doped TMDs, the second neighbor interaction is substantial  and furthermore the strength of the SOC can also be tailored by changing the metal dopant $M^\prime$, so that these materials can serve as a rich laboratory for studying the TLL behavior.

\subsection{Half-Filled Defect Bands: Mott-Hubbard Insulating State and Unusual Band Widening}

\begin{figure*}
\includegraphics[scale=0.65]{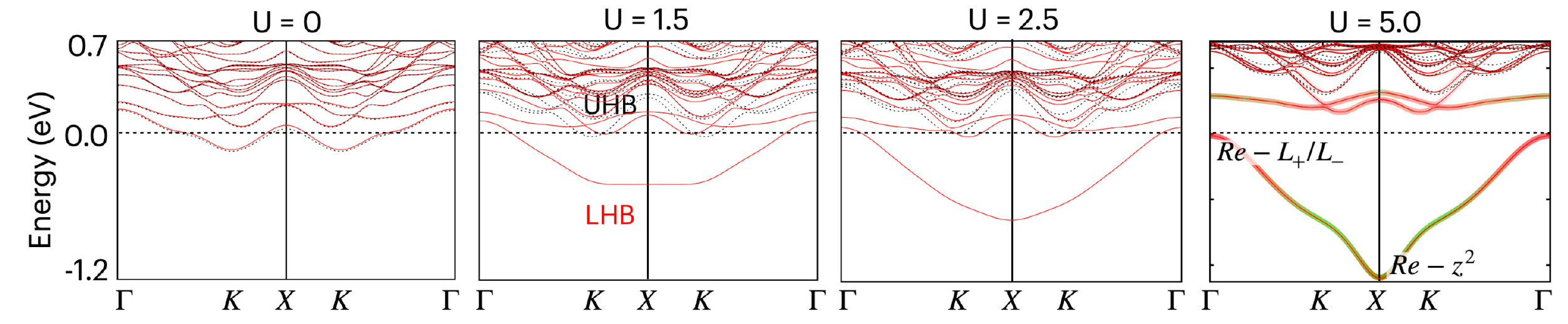}
    \caption{The DFT+$U$ spin polarized band structures of MoSe$_2$/Re. The transition to Mott insulating phase occurs through the formation of lower and upper Hubbard bands (LHB and UHB)). The LHB widens with $U$, which is unusual and is interpreted to be due to its multiple-orbital origin.}
    \label{Fig8}
\end{figure*}
\begin{figure}[h]
\includegraphics[scale=0.165]{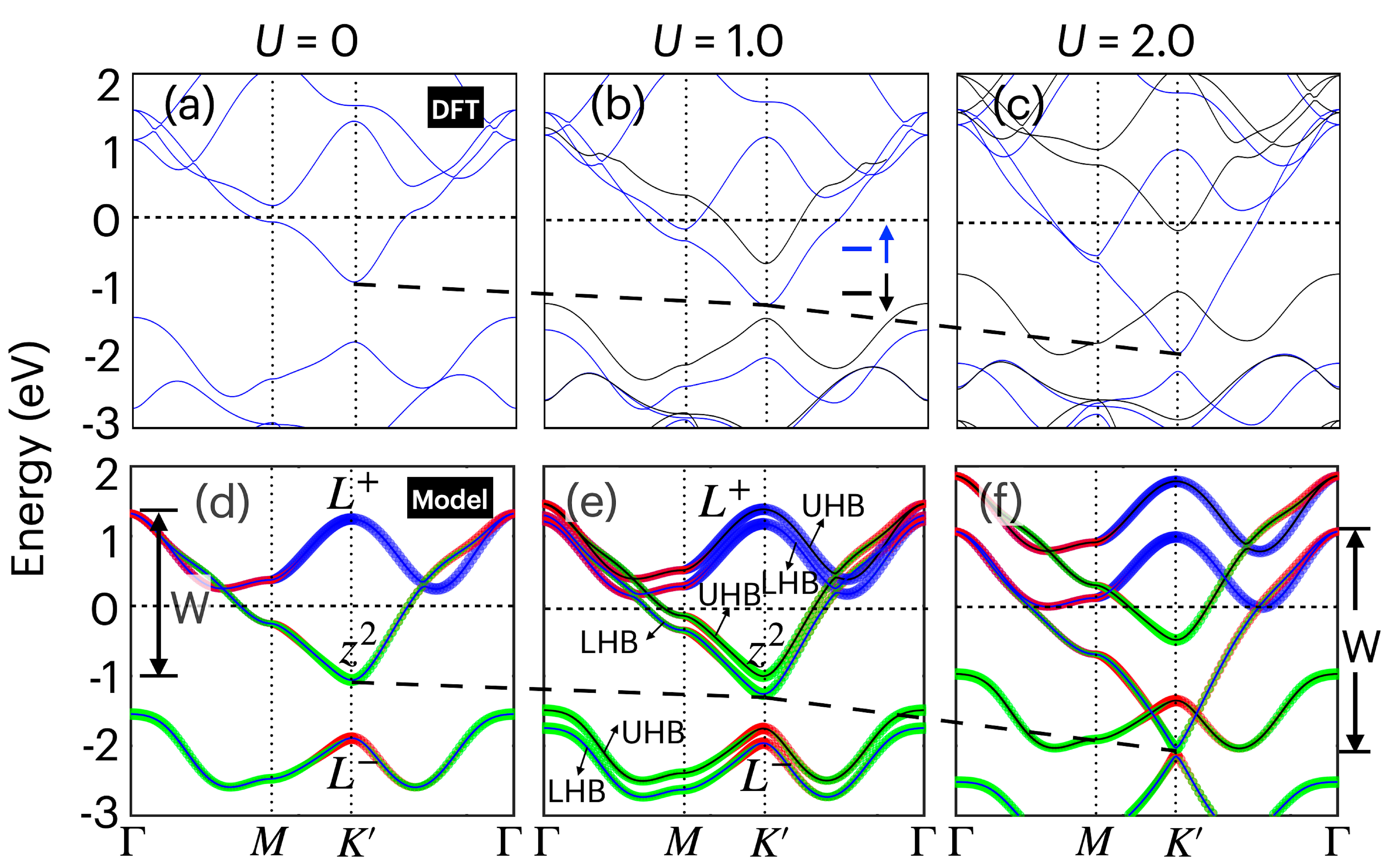}
\caption{The spin-polarized band structure of 2H-ReSe$_2$ monolayer as obtained from DFT+$U$ calculations (upper row) and three-orbital based Hubbard model (lower row) showing excellent agreement. The evolution of the bottom of the LHB with $U$ has been indicated by the dashed line. There is an excellent agreement between the model and DFT results. Like the chain-doped system MoSe$_2$/Re (see Fig. \ref{Fig8}), the LHB bandwidth as indicated in (d) and (f), increases with increasing strength of the onsite Coulomb repulsion.}
\label{Fig9}
\end{figure}
For the half-filled case, the 1D defect bands would not show the TLL behavior due to the presence of a charge gap for all values of $U$ within the Hubbard model. In the strong-coupling limit ($U/W \rightarrow \infty$), which is  the case here, the mean-field treatment adequately describes the gross physics of the system, and this is accomplished by the DFT + $U$ formalism in the density-functional theory. 

We find two surprising features for the 1D defect bands:
(i) Instead of an AFM ground state expected from the Hubbard model physics at half filling, we find that the ground state is usually ferromagnetic.  
(ii) With $U$, the bandwidth of the Hubbard bands increases instead of decreasing which is normally the case for strongly-correlated systems where the motion of the electron is inhibited due to correlated motion, resulting in a larger effective mass. The increase of the bandwidth of the Hubbard bands with $U$ can be explained in terms of a multi-band model as discussed below in some detail.

A plausible explanation for the ferromagnetism is that we don't have an isolated 1D Hubbard chain, rather that the electrons are coupled via the electrons of the host materials. In fact, our charge density contour plot in Fig. \ref{Fig3} infers that in the case of MoTe$_2$/Tc there is a covalent bonding between the 1D states with the nearest-neighbor Te-p states of the host material to favor the ferromagnetic ordering.

The band-widening feature can be explained using a multi-band Hubbard model. In essence,
different parts of the BZ consists of different type of orbitals (for instance, $z^2$ at $K$ and $L^-$ at $\Gamma$ for the defect bands as seen from Fig. \ref{Fig1}), which can shift around in energy depending on the magnitude of $U$. 

To demonstrate the case of Mott insulating phase and band widening, we examine the case of MoSe$_2$/Re as an example. The non-magnetic band structure shown in Fig. \ref{Fig6} indicates that the half-filled 1D bands with the bandwidth of about 0.24 eV (see Table \ref{Table-II}). From spin-polarized DFT + $U$ calculations, we find  (see Fig. \ref{Fig8}) that with increasing $U$, the degenerate non-magnetic half-filled bands spin separate to form a more-occupied lower Hubbard and a less-occupied upper Hubbard band. With further increase in $U$, the two Hubbard bands separate completely to form a gap, leading to a Mott insulating state. We see similar effects for other chain-doped compounds, which are not shown here to avoid redundancy. 

This is in contrast to the  conventional Mott insulator, where increasing $U$ makes the Hubbard bands narrower, as the electron's motion becomes restricted due to electron correlation, resulting in a larger effective mass. In the present case, there is a rapid increase in the bandwidth with an increase in $U$. For example, by increasing $U$ from 0 to 5 eV, the bandwidth has gone from 0.4 eV to 1.1 eV, as can be seen in Fig. \ref{Fig8}.

To describe this physics, we have adopted
 a multi-orbital Hubbard model with three orbital basis ($z^2$, $xy$, and $x^2-y^2$), with the Hamiltonian
 \begin{equation}    \label{MOH}
  \begin{split}
    H & = \sum_{i\mu} \epsilon_{i\mu} c_{i\mu}^\dagger c_{i\mu}+ \sum_{ij;\mu\nu} t_{i\mu j\nu} c_{i\mu}^\dagger c_{j\nu} + h.c. + \\ 
    & U \sum_{i\mu} n_\mathrm{i\mu\uparrow} n_\mathrm{i\mu\downarrow} + (U^\prime -\frac{J_\mathrm{H}}{2}) \sum_{i\mu < \nu} n_\mathrm{i\mu} n_\mathrm{i\nu} \\ 
    & -2J_\mathrm{H} \sum_{i,\mu < \nu} S^z_\mathrm{i\mu} \cdot S^z_\mathrm{i\nu}.
\end{split}
\end{equation}
Here, the first two terms describe the onsite and the kinetic energy of the electrons, while the third and fourth terms are the energy cost of having the electrons in the same or different orbitals at the same lattice site. The last term defines the Hund's rule coupling. The relation $U^\prime$ = $U$ - $2J_\mathrm{H}$ has been used here with J$_H/U$ ratio estimated to be 0.03 based on the agreement between DFT and model band structures. The $n_\mathrm{i\mu}$ = $n_\mathrm{i\mu\uparrow}$ + $n_\mathrm{i\mu\downarrow}$ are the occupation numbers which are obtained from the DFT+$U$  density matrix. 

It is too complex to solve the problem for the supercell of MoSe$_2$/Re. Since the defect bands originate from the doped ReSe$_2$ chain, and the  bulk ReSe$_2$ shows a similar band-widening behavior as well (see Fig. \ref{Fig9}), we study the phenomenon for bulk ReSe$_2$
using the multi-orbital Hubbard model, Eq. (\ref{MOH}).

The results, obtained from both the DFT + U calculations as well as the multi-orbital Hubbard model, are shown in Fig. \ref{Fig9}.  From the top panel in the figure, we observe that for $U$ = 0, it has three spin degenerate bands in the vicinity of the $\epsilon_F$, like in all 2H-TMD compounds. Out of the three, one is completely occupied, another is completely unoccupied and the third one is a half-occupied band. With an increase in $U$, the half-occupied band becomes spin non-degenerate. Also, like the case of MoSe$_2$/Re, the lower Hubbard sub-band becomes more dispersive. The model results, shown in the lower panel of Fig. \ref{Fig9}, match very well 
with the DFT+U results, and the widening of the lower Hubbard band is seen from both results (indicated by the long-dashed lines). 
As mentioned already, the lowest conduction band running through $E_F$ has a combination of the orbital characters $L^{\pm}$ (dominant around  $K^\prime$) and $z^2$ (dominated around $\Gamma$).
The band-widening happens because the inter-orbital interaction term
(the fourth term in the Hamiltonian Eq. \ref{MOH}) lowers the onsite energy for $z^2$ orbital (dominant around $K^\prime$), while increasing the same for  the $L^+$ orbitals (dominant around $\Gamma$). In fact, when this term was switched off, there is a band narrowing, clearly seen for the LHB (see Fig. S12 in SM). 


\section {Summary}
To summarize, by employing density functional theory calculations and theoretical models, we show that the chain-doped transition metal dichalcogenides (MX$_2$/M$^\prime=$M$_{n-1}$M$^\prime_1$X$_{2n}$), with a M$^\prime$ dopant chain along the zigzag direction, form a sharply localized one-dimensional (1D) band structure. While the 1D states are strongly confined along the lateral direction, they are highly mobile along the chain direction. The localization in the lateral direction is interpreted in terms of the bound states of the bare potential of the dopant chain. The partially-filled 1D bands provide a platform to explore exotic spin-orbit coupled one-dimensional quantum phases and properties. These include the Tomonaga Luttinger liquid (TLL) behavior, ferromagnetic Mott insulator, Rashba type spin-orbit coupling and valley-dependent optical transition. The half-filled 1D bands are ideal candidates for stabilizing the anti-ferromagnetic Mott insulating phase. However, the interaction between the 1D states via the host X-p states makes it ferromagnetic and insulating.  When the 1D bands deviate from half-filling, the substantial second-neighbor interactions between the M$^\prime$ states make it favorable for practical realization of the TLL behavior. The deviation from half-filling can be achieved via impurities and gate biasing.

The widening of the lower Hubbard subband with increasing onsite Coulomb repulsion strength in these chain-doped systems is a non-trivial outcome of this study. This phenomenon, which is anti-intuitive and goes against the conventional assumption of band narrowing with increasing repulsion strength, has hardly been observed in the literature, which makes this class of materials worth exploring for non-trivial quantum transport and phases. We have explained the cause of band widening by developing a multi-orbital Hubbard model. Another important outcome of the present study is that due to the presence of an intrinsic electric field along the lateral direction, the 1D bands are Rashba spin-split and provide a new mechanism to tune the valley-dependent optical transition in MX$_2$/M$^\prime$.

Our study opens new avenues to tailor 1D quantum physics in 2D TMDs. With the advent of state-of-the-art techniques such as low-energy ion implantation, dislocation climb mechanism, etc., the chain-doped TMDs can be synthesized in a controlled manner. In addition to the electron doping and emergent properties as discussed in this work, hole doping is also equally likely to introduce interesting features such as orbital and spin Hall effects in such chain-doped compounds. As a whole, we believe that the present study will excite experimenters and theoreticians alike to envisage exotic quantum phenomena and applications.

\textit{Acknowledgements:} This work is funded by the Department of Science and Technology, India, through Grant No. CRG/2020/004330. SS thanks SERB India for the VAJRA fellowship. BRKN acknowledges the support of HPCE, IIT Madras for providing computational facilities.

%


\renewcommand{\thetable}{S\arabic{table}}
\renewcommand{\thefigure}{S\arabic{figure}}

\widetext

\vspace{2cm}
\section*{\large Supplementary material}	
\vspace{1cm}	

\section*{I. DFT fitted TB band structures of 1D localized Tc bands in WTe$_2$/Tc.}
\begin{figure}[h]
    \centering
    \includegraphics[scale=0.2]{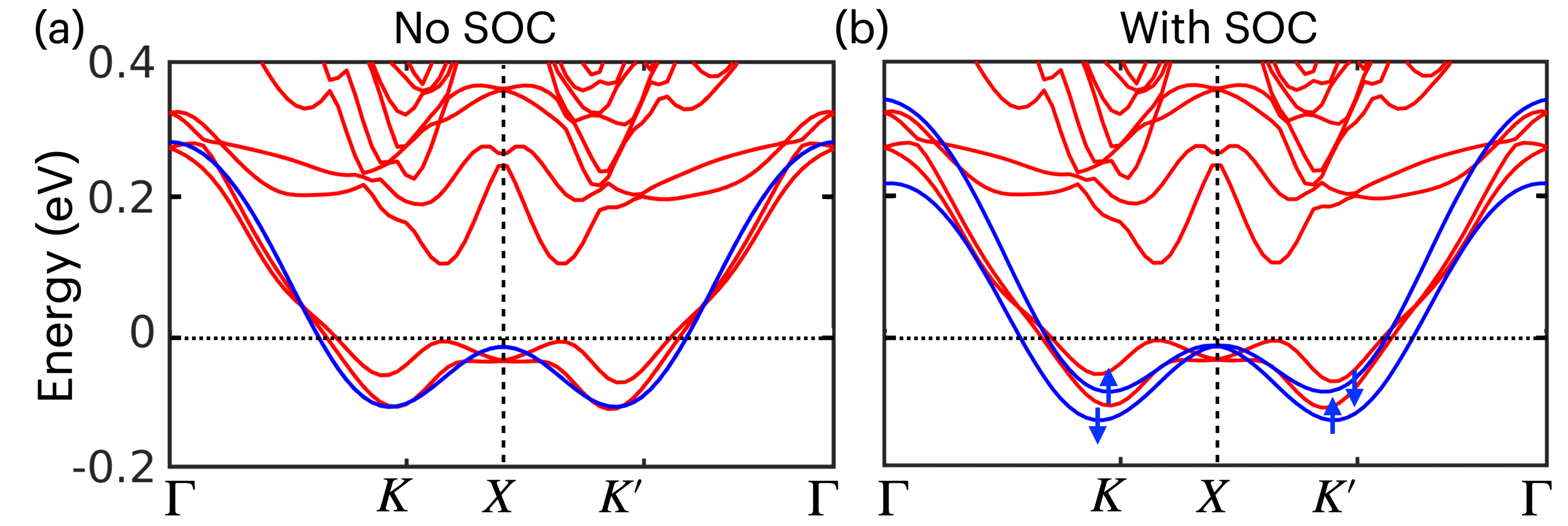}
    \caption{(a) DFT (red) and TB band structures (blue) of WTe$_2$/Tc. (b) The spin texture of the TB bands with an electric field of strength 0.2 eV.}
    \label{fig:tb}
\end{figure}

To estimate the hopping strengths for the dopant chain, we have designed a single-band tight-binding (TB) model including hopping interactions upto third-neighbor. Furthermore, to obtain spin texture, the electric field is applied perpendicular to the chain along the y-direction. The Hamiltonian is provided in Eq. 2 of the main text. Taking WTe$_2$/Tc as an example, we have shown the TB band fitting with the DFT bands (see Fig. \ref{fig:tb}). The estimated hopping strengths are listed in Table I of the main text.
\section*{II. Orbital resolved band structure of WTe$_2$/Tc}
\begin{figure}[h]
    \centering
    \includegraphics[scale=0.2]{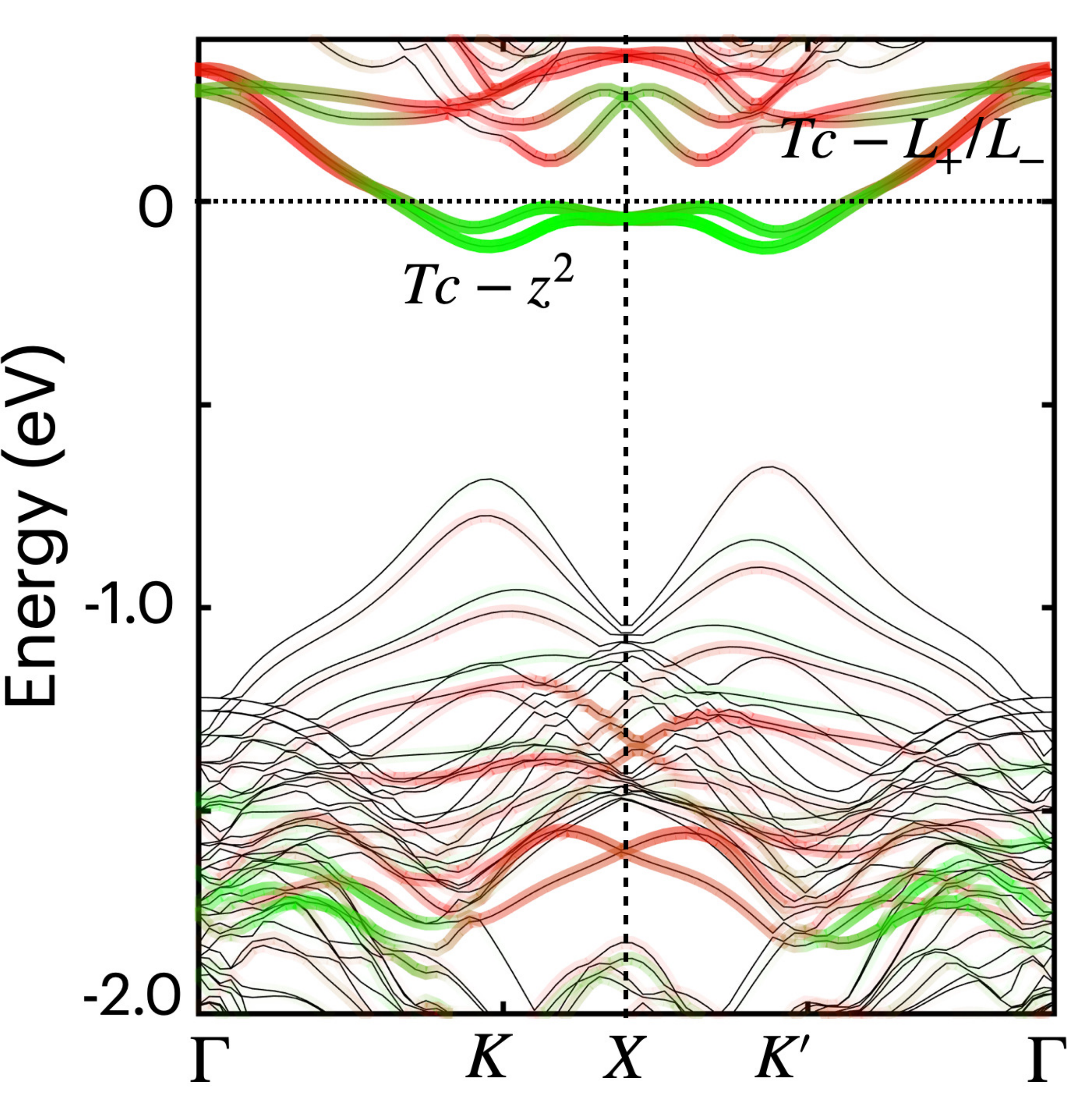}
    \caption{The orbital and atom resolved band structure of WTe$_2$/Tc}
    \label{fig:WTe2/Tc}
\end{figure}

Figure \ref{fig:WTe2/Tc} shows the density functional theory (DFT) obtained orbital resolved band structure of Tc atom in WTe$_2$/Tc. For the half-filled 1D defect band, while the valleys are dominated by $z^2$ orbitals, rest band in the Brillouin zone is dominated by $L_\pm$ orbitals. As the figure shows, the valleys of unoccupied Tc bands are formed by $L_\pm$ orbitals. Therefore, the presence of z$^2$ ($l$=0) and $L_\pm$ ($l$=$\pm$1) states at the valley points of the partially occupied and unoccupied defect bands opens up the possibilities of valley-dependent optical transitions. 
\section*{III. Widening of the Hubbard band}
\begin{figure}[!h]
    \centering
    \includegraphics[scale=0.3]{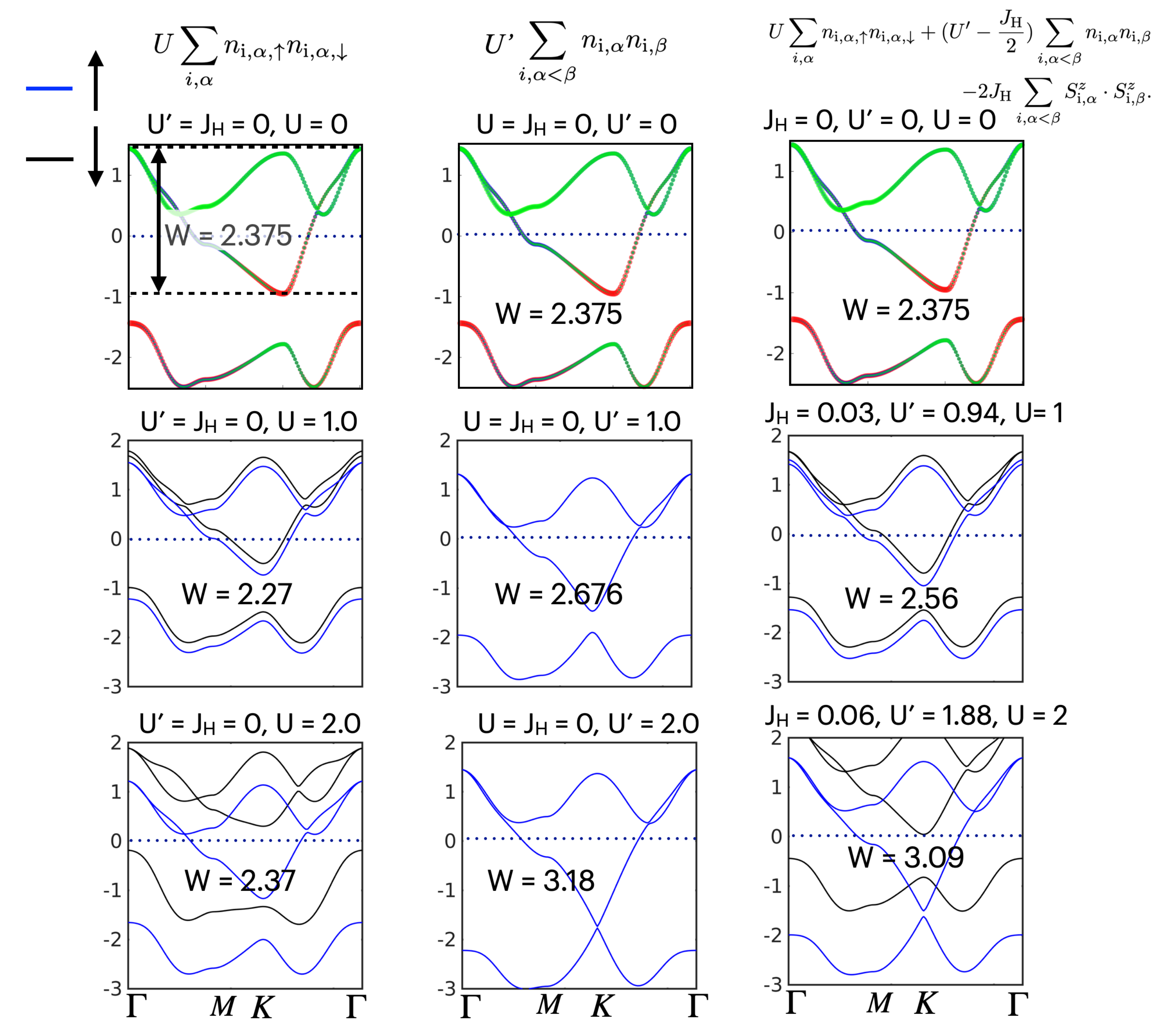}
    \caption{The modeled band structures of pristine ReSe$_2$ obtained by considering only the onsite intra-orbital Coulomb repulsion (first column), only inter-orbital Coulomb repulsion (second column), and inter-orbital, intra-orbital, and Hund's coupling (third column), respectively. The relation $U^\prime = U - 2J_H$ between Kanamori parameters has been used here. More or less the bandwidth remains same with varying onsite Coulomb repulsion $U$, however, it changes significantly when only inter-orbital interaction is switched on. The bandwidths obtained in third column closely resembles with those in second column as the Hund's coupling is weak (J$_H$/U=0.03) and $U$ hardly affects the bandwidth. Hence, inter-orbital Coulomb interaction $U^\prime$ plays the dominant role in band widening (see main text).}
    \label{fig:S3}
\end{figure}

To investigate the underlying interaction causing the band widening, see subsection-IV(D) of the main text, we employ the multi-orbital Hubbard model for pristine ReSe$_2$. The obtained modeled band structures with varying $U$ strength are shown in Fig. \ref{fig:S3}. The band structures presented in the first, second, and third columns are obtained by considering the intra-orbital Coulomb repulsion (the third term of $H$), inter-orbital Coulomb repulsion (the fourth term of $H$), and third, fourth, and fifth terms in $H$, respectively. As inferred from the first column, the bandwidth remains almost unchanged with varying intra-orbital interaction, however, with inter-orbital interaction, the bandwidth increases substantially from 2.37 to 3.18 eV. With the inclusion of all terms in $H$, more or less, the bandwidth remains unchanged as compared to the case when only inter-orbital interaction is considered. This is expected as Hund's coupling strength is weak (J$_H$/U=0.03), and $U$ hardly affects the bandwidth. This clearly depicts the dominant role of inter-orbital Coulomb repulsion in band widening.  
\begin{eqnarray}
    H_\mathrm{int} = \sum_{i,\mu} \epsilon_{i\mu} c_{i\mu}^\dagger c_{i\mu}+ \sum_{ij;\mu\nu} t_{i\mu j\nu} c_{i\mu}^\dagger c_{j\nu} + h.c. + 
    U \sum_{i,\mu} n_\mathrm{i,\mu,\uparrow} n_\mathrm{i,\mu,\downarrow} + (U^\prime -\frac{J_\mathrm{H}}{2}) \sum_{i,\mu < \nu} n_\mathrm{i,\mu} n_\mathrm{i,\nu}
    -2J_\mathrm{H} \sum_{i,\mu < \nu} S^z_\mathrm{i,\mu} \cdot S^z_\mathrm{i,\nu}.\nonumber
\end{eqnarray}

\end{document}